\documentclass[nofootinbib,prl,twocolumn,amsmath,amssymb,aps]{revtex4-1}
\usepackage{graphicx}
\usepackage{hyperref}

\usepackage{float}
\hypersetup{
     colorlinks   = true,
     linkcolor    = blue,
     citecolor    = blue
}
\usepackage[all]{hypcap}

\begin{document}

\title{Ferromagnetism and its stability from the one-magnon spectrum in twisted bilayer graphene}
\author{Yahya Alavirad$^1$}
\author{Jay Sau$^1$}

\affiliation{$^1$Department of Physics, Condensed Matter theory center and the Joint Quantum Institute, University of Maryland, College Park, MD 20742, USA}

\date{\today}

\begin{abstract}
We study ferromagnetism and its stability in twisted bilayer graphene. We work with a Hubbard-like interaction that corresponds to the screened Coulomb interaction in a well-defined limit where the Thomas-Fermi screening length $l_\text{TF}$ is much larger than monolayer graphene's lattice spacing $l_g \ll l_\text{TF}$ and much smaller than the Moir\'e super lattice's spacing $ l_\text{TF} \ll l_{\text{Moir\'e}}$. We show that in the perfectly flat band ``chiral" limit and at filling fractions $\pm 3/4$, the saturated ferromagnetic (spin and valley polarized) states are ideal ground states candidates in the large band-gap limit. By assuming a large enough substrate (hBN) induced sub-lattice potential, the same argument can be applied to filling fractions $\pm 1/4$. We estimate the regime of stability of the ferromagnetic phase around the chiral limit by studying the exactly calculated spectrum of one-magnon excitations. The instability of the ferromagnetic state is signaled by a negative magnon excitation energy. This approach allows us to deform the results of the idealized chiral model (by increasing the bandwidth and/or modified interactions) towards more realistic systems. Furthermore, we use the low energy part of the exact one-magnon spectrum to calculate the spin-stiffness of the Goldstone modes throughout the ferromagnetic phase. The calculated value of spin-stiffness can determine the excitation energy of charged skyrmions.
\end{abstract}
\maketitle

\emph{Introduction}- Ferromagnetism is the most familiar form of magnetic order. Despite the long history of ferromagnetism, most of our current understanding is based on simple Hartree-Fock (mean-field) calculations\cite{mahan}. These calculations are known to greatly overestimate the ferromagnetic tendency of electronic systems. Several improvements over the Hartree-Fock method have been proposed\cite{kanamori,stohr}. Yet, the overall progress in this direction, has not lead to a theory that provides reliable diagnostics for which systems would be ferromagnetic.

A practically useful guide is provided by the Hund's rule that predicts ferromagnetic spin-polarization in partially filled degenerate sets of energy states (orbitals). Specifically, the exchange term in the Coulomb interaction reduces the Coulomb repulsion between electrons of similar spin favoring spins to align with each other. Interestingly, the same general principle appears to apply to quantum Hall ferromagnetism. In both of these cases, the degeneracy of the non-interacting energy eigenstates seems essential to enhance the effect of the ferromagnetic exchange.  

While a faithful treatment of magnetism in electronic systems is complicated, the limit of strong on-site Coulomb interaction $U$, the so-called Hubbard interaction\cite{hubbard}, has been demonstrated to lead to anti-ferromagnetic Ne\'el order on an energy-scale proportional to Heisenberg super-exchange\cite{mahan}. The magnetic order has been shown to flip to ferromagnetism in the limit of exactly one-hole and infinite onsite repulsion $U$\cite{Nagaoka,Thouless}. These results were extended by Lieb to the half-filled Hubbard model with an imbalance in the number of sub-lattices\cite{lieb}, establishing the possibility of itinerant ferromagnetism. These ideas of enhancement of magnetism by local interaction and of ferromagnetism by degeneracy of non-interacting states were later shown to reinforce each other through the demonstration of ferromagnetism in half-filled lattice models with Hubbard interactions that have a degenerate manifold of states in the form of a flat band\cite{tasaki,mt1,mt2,mt2.5,mt3,mt4,mt5,mt6,mt7,mt8,mt9,mt10,mt11}. The latter class of results constitute what's usually called ``flat band ferromagnetism".

Despite the large variety of theoretical models demonstrating spontaneous ferromagnetism as well as competing magnetic and itinerant phases, physical realizations of such models are lacking. Recent experimental breakthroughs in the area of multi-layer graphene both in the quantum Hall regime and without magnetic fields provide hope for the realization of such systems. In the quantum Hall regime, graphene provides the opportunity to break the flatness of a Landau level by introducing a lattice potential on the scale of a magnetic length that has been shown to create a Hofstadter spectrum\cite{Hunt1427}. Based on arguments in the previous paragraphs, one expects such a broadening of the Landau levels to compete with quantum Hall ferromagnetism in an interesting way. The latter case of multi-layer graphene without a magnetic field is a more unexpected direction and appears to have evidence for ferromagnetism. More specifically, large peaks in density of states associated with nearly flat-bands and concomitant appearance of correlated phenomena have recently been observed in twisted bilayer, twisted double-bilayer, and ABC trilayer graphene\cite{tbg1,tbg2,tbg3,tbg4,tbg5,tbg6,tbg7,tbg8,tbg9,tbg10,tbg12,tbg11,tbg13,tbg14,young}. Some of these systems have also shown evidence for ferromagnetism\cite{tbg6,tbg8,tbg9,tbg10,tbg12,tbg14,young} near the ``flat-band" limit. Additionally, the possibility of tuning these systems out of the ``flat-band" regime suggests the fascinating possibility of studying multiple phases and transitions between them of insulating and itinerant magnetic systems.

In this work, we consider the particular example of twisted-bilayer graphene(TBLG). We start by focusing on the so-called ``chiral" limit of the realistic models for TBLG, where the spectrum supports a band that is exactly flat~\cite{vf1,vf2,vf3}. We work with particular form of Hubbard interaction that we argue can emerge from Thomas-Fermi 
screening of the Coulomb interaction. We then show that in this limit and at filling fractions $\pm 3/4$, the saturated spin and valley polarized states are ideal ground states candidates of the system. By assuming a large enough substrate (hBN) induced sub-lattice potential, the same argument can be shown to hold for filling fractions $\pm 1/4$. The topology of the TBLG band structure, guarantees that all the ferromagnetic states discussed above are also associated with a quantized anomalous Hall response\cite{senthil,zaletel}. We study the local stability of the ferromagnetic phase around the chiral limit by studying the \textit{exactly} calculated spectrum of one-magnon excitations. The instability of the ferromagnetic state is signaled by a negative magnon excitation energy. This approach allows us to deform the results from the idealized chiral model (by increasing the bandwidth and/or modified interactions) towards results for more realistic systems. We use the low energy part of the exact one-magnon spectrum to predict the spin-stiffness of the Goldstone modes in the ferromagnetic phase as the realistic system is approached. The effect of spin-stiffness can be potentially determined from skyrmion-induced transport phenomena.

\emph{Band structure of TBLG}\label{etbg}- TBLG corresponds to two layers of graphene stacked on top of each other with a relative twist angle $\theta$. For small twist angles $\theta$, and within the leading harmonic approximation, this system forms a periodic pattern called a Moir\'e pattern. In this limit, the noninteracting physics can be well approximated by the Bistritzer and Macdonald continuum model\cite{bm,bm2}. Following the notation of Ref.~\cite{koshino}, the dimensionless single valley ($\zeta=\pm1$ is the valley index) Bistritzer and Macdonald Hamiltonian can be written in the layer $(1,2)$ and sub-lattice  $(A,B)$ basis $(A_1,B_1,A_2,B_2)$ as,
\begin{align}
H_{\text{BM}}^\zeta= \left(\begin{array}{cc} H_1 & U^\dagger(\mathbf{r}) \\ U(\mathbf{r}) & H_2  \end{array}\right),
\end{align}
where,
\begin{align}
H_l= - R(\pm \theta/2) (k-K_\zeta^l) .(\zeta \sigma_x,\sigma_y) + \Delta_l \sigma_z
\end{align}
and,
\begin{align}\label{ut}
U(\mathbf{r})=&\left(\begin{array}{cc} \alpha_0 & \alpha_1 \\ \alpha_1 & \alpha_0 \end{array}\right)+\left(\begin{array}{cc} \alpha_0 &  \alpha_1 e^{-2\pi i \zeta/3} \\ \alpha_1 e^{2\pi i \zeta/3} & \alpha_0 \end{array}\right) e^{i\zeta G_1.r}   \\ \newline \nonumber
&+\left(\begin{array}{cc} \alpha_0 &  \alpha_1 e^{2\pi i \zeta/3} \\ \alpha_1 e^{-2\pi i \zeta/3} & \alpha_0 \end{array}\right) e^{i\zeta (G_1+G_2).r}.
\end{align}
Here, $G$'s are the reciprocal lattice vectors of the Moir\'e lattice, and $K_\zeta^l$'s are the location of monolayer Dirac points in the Brillouin zone. $\Delta_l$'s are the (hBN) induced sub-lattice potentials. In monolayer graphene $\Delta$ is known to be able to reach around $\Delta\approx 0.1-0.15$ (in dimensionless units used here or equivalently $15-30$ meV)\cite{hbn1,hbn2,hbn3}.

The dimensionless parameters $\alpha_0,\alpha_1$ are given by,
\begin{align}
\alpha_{0}=\frac{3 w_{AB} a_0}{8 \pi v_0 \sin (\theta/2)} ~; ~ \alpha_{1}=\alpha_{0} \frac{w_{AA} }{w_{AB}}.\label{eqalpha}
\end{align}
$a_0$ and $v_0$ are respectively, the monolayer graphene's lattice spacing and the Fermi velocity. $w_{AB}$ and $w_{AA}$ are roughly, the hopping amplitudes in the $AA$ and $AB/BA$ stacking regions. In the realistic system $\alpha_0$ is expected to be around $\alpha_0\approx0.586$ and $ \frac{w_{AA} }{w_{AB}}$ is expected to be around $ \frac{w_{AA} }{w_{AB}}\approx 0.8$\cite{koshino}. Many interesting features related to this non-interacting model has been extensively studied in the past year\cite{ni1,ni2,ni3,ni4,ni5,ni6,ni7,ni8,1907.02856,PhysRevB.99.165112}.

\emph{Interaction model for TBLG}- In this letter, we mostly work with a simple yet reasonable model for interactions in TBLG. The effect of more general interactions is discussed later.

 We start by considering the RPA screened Coulomb interaction $V(q)$. The exact $V(q)$ has been found to be rather complicated\cite{thomale}. An approximation for the small $q$ behavior of $V(q)$ can be obtained from simple Thomas-Fermi screening arguments. A rough estimate for the Thomas-Fermi screening wave-vector $q_\text{TF}$ can be obtained from the monolayer-graphene results of Ref.\onlinecite{sankar}  (by using a renormalized Fermi velocity). This results suggest that $G_{\textit{Moir\'e}}\ll q_\text{TF} \ll G_{\text{graphene}}$. Since $q_\text{TF} \ll G_{\text{graphene}}$, in this regime the interaction is independent of  layer and sub-lattice separation (these distances are much smaller than $1/q_\text{TF} $). Also the rotation angle $\theta$ is small, therefore it's effect on inter-particle distance can be dropped (interaction becomes layer independent). Note that the low energy states included in the continuum model of Bistritzer and Macdonald are only the states close to Dirac points $|k-K|< \mathcal{O}(1) G_{\textit{Moir\'e}}$. Since  $G_{\textit{Moir\'e}}\ll q_\text{TF}$, in this regime $V(q)$ is effectively constant.  Similarly, because $q_\text{TF} \ll G_{\text{graphene}}$, the inter-valley scattering terms are strongly suppressed and the valley index becomes an effective good quantum number (approximate $U(1)_v$ symmetry). Putting everything together leads to the following simplified form of the interaction,
\begin{align}\label{hint}
V=U \sum_q  \sum_{(\sigma',v',s',l')\neq (\sigma,v,s,l)} \rho_{\sigma,v,s,l} (q) \rho_{\sigma',v',s',l'} (-q).
\end{align}
Here $\rho_{\sigma,v,s,l}(q)=\sum_k c^\dagger_{\sigma,v,s,l,k+q} c_{\sigma,v,s,l, k}$ is the density wave operator. $\sigma,v,s,l$ are the spin, valley, sub-lattice, and layer indices respectively. The term with all the indices equal is dropped since it only renormalizes the chemical potential (which is irrelevant at a fixed filling). 

\begin{figure}[t]
\centering
	\vspace{1mm}
\includegraphics[width=\columnwidth,keepaspectratio]{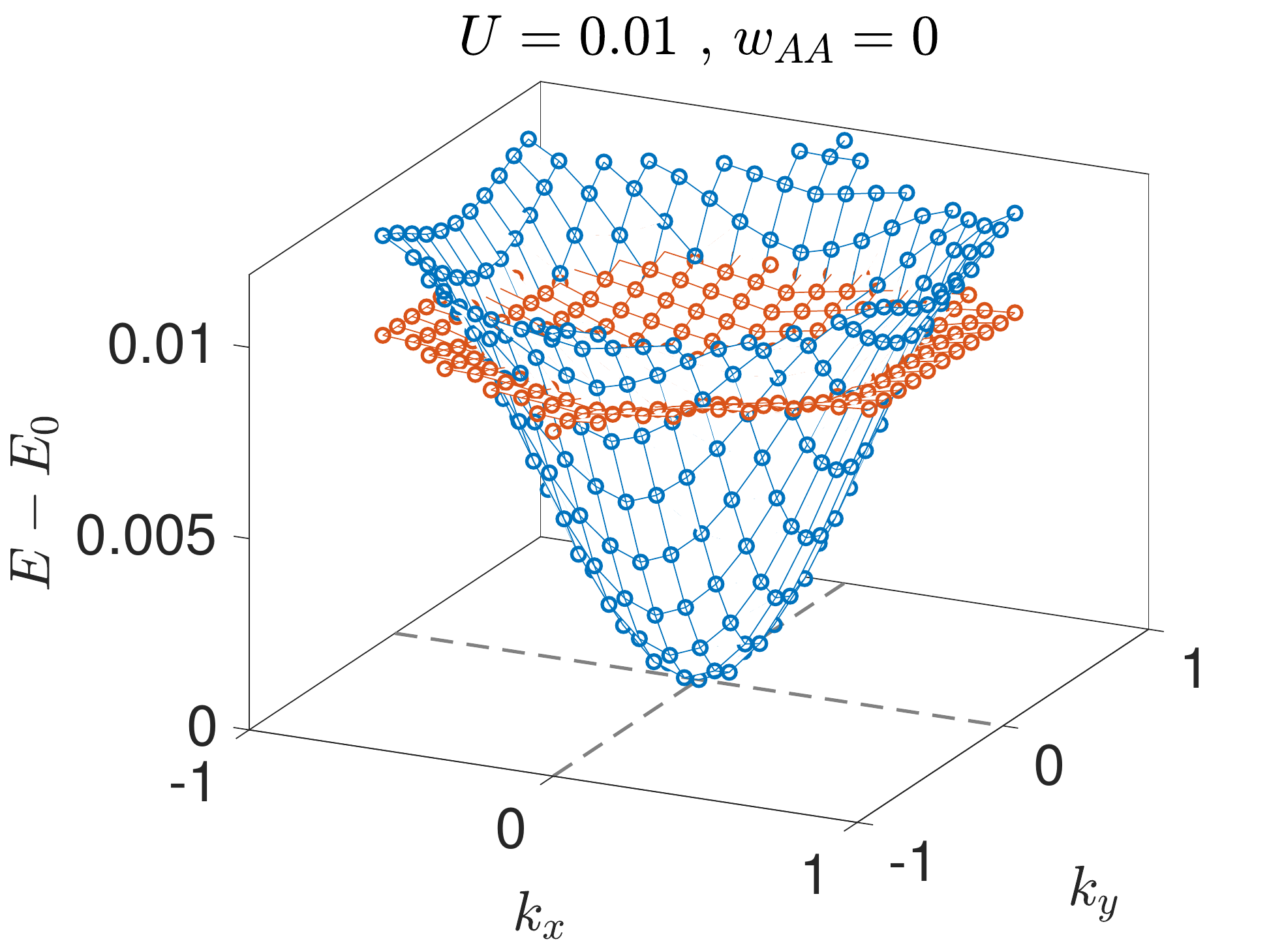} 
\caption{Lowest one-magnon band spectrum of the TBLG in the chiral limit. Energies are measured with respect to the fully polarized state. The gapless blue curve corresponds to the single-spin flip branch associated with the $SU(2)$ breaking Goldstone mode. The gapped red curve corresponds to the single-valley flip branch associated with breaking the discrete time-reversal symmetry. Energies are in units of $\frac{8 \pi v_0 \sin (\theta/2)}{3  a_0}\approx 0.19 \text{eV}$. $k_x$ and $k_y$ are in units of $\frac{8 \pi \sin (\theta/2)}{3  a_0}$. We have used the interaction form $V$ (Eq.\eqref{hint}).}
\label{2.pdf}
\end{figure}

\emph{Ferromagnetism in the perfectly flat band limit of TBLG}- Let us now consider the effect of the interaction Eq.~\ref{hint} on TBLG in the chiral limit ($\alpha_0=0.586$ and $\alpha_1=0$ in Eq.~\eqref{ut}).  In this limit, the flat band wave-functions can be taken to be sub-lattice polarized~\cite{vf1} so that the sub-lattice index $s$ is a good quantum number in addition to $\sigma,v$. We now have $8$ degenerate flat bands bands that can be labeled by spin, valley and sub-lattice indices $\sigma,v,s$.  

Assuming the interaction parameter $U$ is small compared to the band gap $W$, we can consider an effective Hamiltonian,
\begin{align}\label{hint2}
H_t=U \sum_{q,l,l',f\neq f'}  P_0\rho_{f,l} (q) P_0 \rho_{f',l'} (-q)P_0,
\end{align}
where $P_0$ is the projection operator into the flat bands 
and $f$ refers to the collective $\sigma,v,s,l$. 
This Hamiltonian (that we focus on here) differs from Eq.\eqref{hint} by ``intra-flavor inter-layer" terms. 
Later, we will show numerically that these terms do not have a significant effect on 
the ground state. 
 The projected density operators 
$P_0\rho_{f,l} (q) P_0$ in $H_t$ commutes with the kinetic energy term in the flat-band limit, 
so that we can ignore the kinetic energy.
 A  spin, valley and sub-lattice polarized state corresponding to fully filling one of these bands labeled by $|f=f_0\rangle=\prod_{k\in\text{MBZ}} c^\dagger_{f_0,k}|0\rangle$ 
is a null state (i.e. zero-energy eigenstate) of $H_t$. To see this note that 
\begin{align}
&P_0\rho_{f,l} (q) P_0|f=f_0\rangle=\sum_{G_\text{Moir\'e}}\Lambda_l(G_\text{Moir\'e}) \delta_{f,f_0}\delta_{q,G_\text{Moir\'e}}|f=f_0\rangle,
\end{align}
which implies that 
\begin{align}
&H_t|f=f_0\rangle=U \sum_{G_\text{Moir\'e},G'_\text{Moir\'e},l,l',f\neq f'}\nonumber\\
&\Lambda_l(G_\text{Moir\'e})\Lambda_{l'}(G'_\text{Moir\'e})\delta_{f,f_0}\delta_{f',f_0}|f=f_0\rangle=0.
\end{align}
Note that the Hamiltonian in $H_t$ in Eq.~\ref{hint2} is non-negative i.e. $\langle H_t\rangle \ge 0$. This becomes manifest if we Fourier transform back into ``real" space,
\begin{align}
&H_t=U \int d^2\bm r \sum_{l,l',f\neq f'}  (P_0 n_{f,l} (\bm r) P_0) (P_0 n_{f',l'} (\bm  r)P_0),
\end{align}
where  $n_{f,l}(\bm r)$ is the real-space density operator. The two parts of each product term commute as they are associated with different values of $f$. They are also both non-negative as they are projected non-negative operators.

Therefore, the null state $|f=f_0\rangle$ is an exact ground state of $H_t$ at filling is $-3/4$ (one electron per unit cell) of the flat band manifold.  Since the chiral limit Hamiltonian is particle-hole symmetric, the same results also hold for the opposite filling fraction $+3/4$. That is, the fully polarized state is an exact ground state at fillings $\pm 3/4$. By assuming large enough (substrate induced) sub-lattice potential $\Delta_t=\Delta_b > U$, the same result can be easily generalized to filling fractions $\pm 1/4$.  Note that the band structure properties of TBLG guarantees that all of the ferromagnetic states discussed here are also associated with a quantized anomalous Hall response\cite{senthil,zaletel,senthil2}.
Within our formalism, we cannot find a justification for considering ferromagnetic states at $\pm1/2$ or $0$(charge-neutrality) fillings. In fact, we believe it is likely that the true ground state of the model at these fillings is not ferromagnetic. However, the mean field studies of Refs.\onlinecite{mf1,mf2,1906.07302}, find that the spin/valley polarized states are good ground state candidates even at $\pm1/2,0$ filling.

\emph{Spin-stiffness and the stability of ferromagnetism in TBLG.}- We now turn to discussing the stability of this ferromagnetic state using the one-magnon spectrum. This is a crucial step as it provides a non-trivial consistency check and allows us to generalize the results of the idealized model (Eq.~\ref{hint2}) to more realistic systems.  If the system is truly ferromagnetic, it is necessary but not sufficient for the $q=0$ state to have the minimum energy (since it is related to the fully polarized state by a $SU(2)$ rotation). This establishes the ferromagnetic state as the local energy minimum. We note that even though in principle the local stability of ferromagnetic state is not enough to guarantee global stability, application of our method to a few known examples (in the supplementary material~\ref{sm}) suggests that in practice the ferromagnetic region of the phase diagram can be identified effectively. We can further use the one-magnon band spectrum of the system to obtain useful information like spin-stiffness.

\begin{figure}[t]
\centering
	\vspace{1mm}
\includegraphics[width=\columnwidth,keepaspectratio]{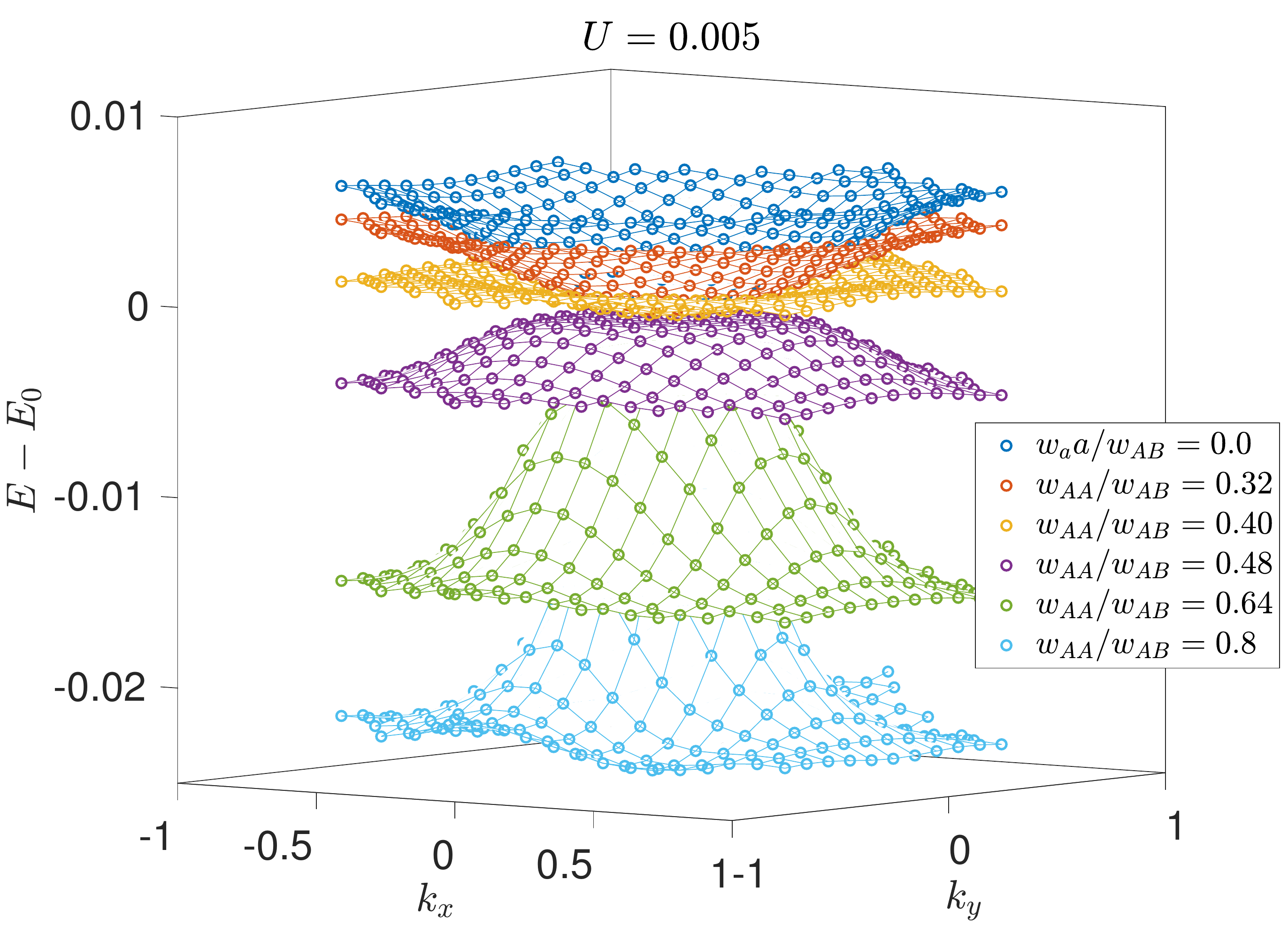} 
\caption{Lowest one-magnon (single spin-flip) band spectrum of the TBLG as the realistic system is approached. Energies are measured with respect to the fully polarized state in the chiral limit. Energies are in units of $\frac{8 \pi v_0 \sin (\theta/2)}{3  a_0}\approx 0.19 \text{eV}$. Here $\Delta_t=\Delta_b=0.1\approx 18meV$. $k_x$ and $k_y$ are in units of $\frac{8 \pi \sin (\theta/2)}{3  a_0}$.We have used the interaction form $V$ (Eq.\eqref{hint}).}
\label{1.pdf}
\end{figure}

A combination translation invariance and flavor conservation ensures that the one-magnon (single-spin or valley flip) Hilbert space is small enough to be accessible using exact diagonalization. We use the \textit{exactly} calculated one-magnon spectrum to study the stability of the ferromagnetic state.
The exactly calculated band-structure of the one-magnon excitations using the interaction $V$ (Eq.~\ref{hint}) is shown in Fig.\ref{2.pdf}. The blue curve corresponds to the single-spin flip branch of excitations associated with the $SU(2)$ breaking Goldstone mode. The red curve corresponds to the single-valley flip branch of excitations associated with breaking of the time-reversal symmetry. Importantly note that since time-reversal symmetry is discrete, the single-valley flip excitations are gapped. As shown in Fig.~\ref{2.pdf} the ferromagnetic state is stable in this case.

\begin{figure}[t]
\centering
	\vspace{1mm}
\includegraphics[width=\columnwidth,keepaspectratio]{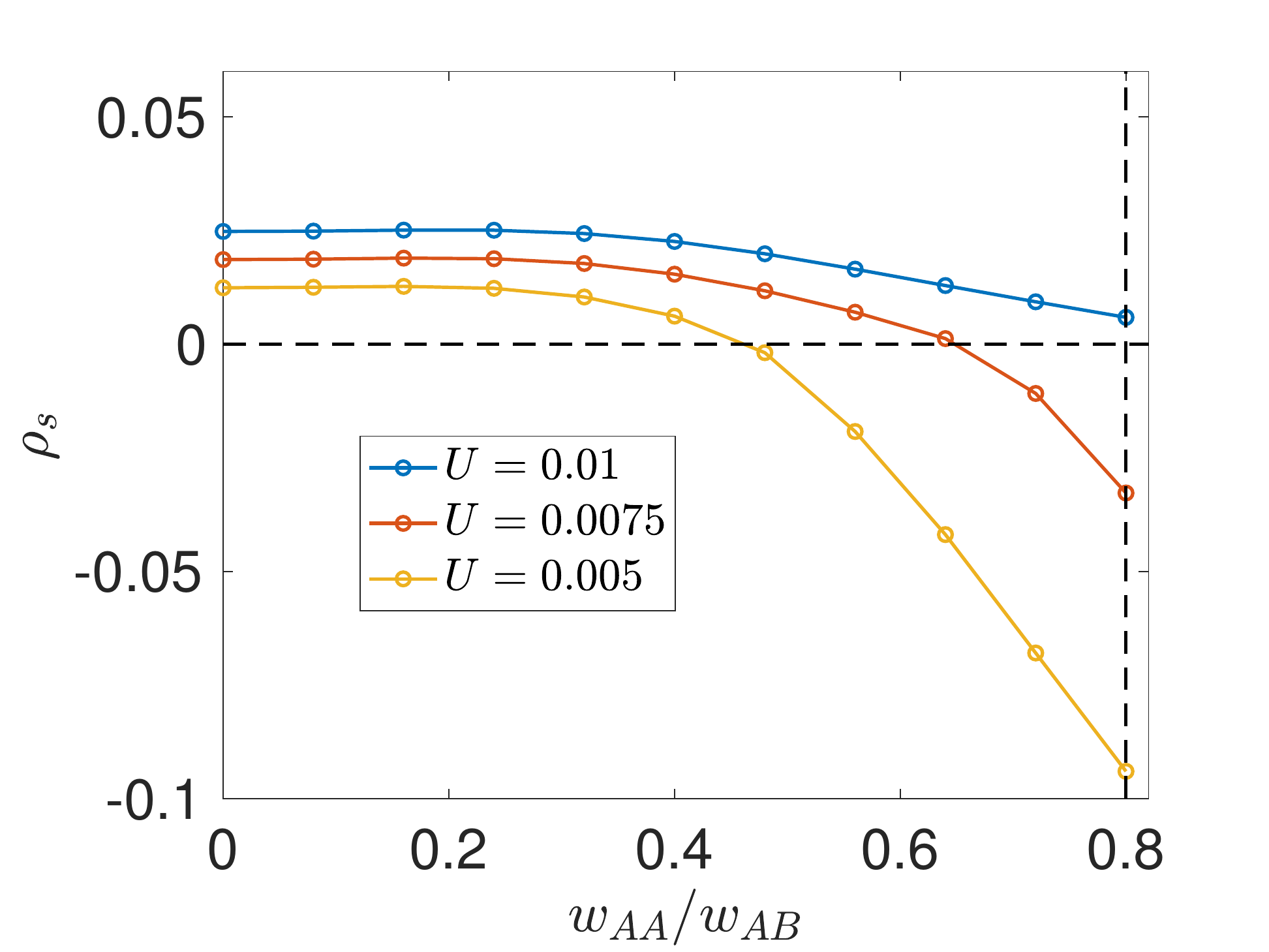} 
\caption{Spin-stiffness associated with the one-magnon band spectrum of the TBLG. A negative spin-stiffness signals the instability of the ferromagnetic state. We have used a fit of the form $E=\rho_s |k|^2$. We have used the interaction form $V$ (Eq.\eqref{hint}).}
\label{3.pdf}
\end{figure}

The numerical stability of the ferromagnetic state shows that the  ``intra-flavor inter-layer" terms
\begin{align}
 \Delta H_t=  2 U \sum_q  \sum_{f}  P_0\rho_{f,l=1} (q)P_0 \rho_{f,l=2} (-q)P_0.
\end{align}
that were ignored in the analytic arguments based on Eq.~\ref{hint2} does not significantly affect the ferromagnetism (we have checked that this term only cause small corrections to the one-magnon spectrum). This term favors layer polarization. However, even approximately layer polarized states do not exist in the flat band subspace (layer polarization is largely fixed by the non-interacting flat-band wave functions). Additionally, such terms constitute a combinatorially small fraction $1/28$ of the total terms of Eq.\eqref{hint}. Adding this term to Eq.\eqref{hint2} can thought of as deviating the ideal interaction form of Eq.\eqref{hint2} towards the more realistic interaction in Eq.\eqref{hint}.

 We now deviate from the chiral limit and proceed to study the stability of ferromagnetic state as the realistic system is approached. For simplicity, here we assume a substrate induced sub-lattice potential $\Delta_t=\Delta_b=0.1\approx 18meV$. This assumption gaps one of the flat bands and helps with the computational complexity. 
We numerically calculate the one-magnon (single spin-flip) spectrum as we approach the realistic parameters $w_{AA}/w_{AB}=0.8$\cite{koshino}. The instability of the ferromagnetic state is identified by a sign-change of the spin-stiffness, or more precisely by looking for one-magnon states whose energy is lower than the ferromagnetic state. Sample results are shown in Fig.\ref{1.pdf}. In Fig.\ref{1.pdf} we have set $U=0.005\approx 1meV$. As shown in the figure, as the realistic parameters are approached, the ferromagnetic state becomes unstable. To study this transition more carefully, we have plotted the calculated value of spin-stiffness $\rho_s$ as a function of $w_{AA}/w_{AB}$ for three different values of $U$ in Fig.\ref{3.pdf}. $\rho_s$ is extracted assuming $E=\rho_s |k|^2$, note that as the instability is approached, the spectrum sometimes does not admit a good quadratic fit. Still, the extracted value can be used to see the general trend. As shown in Fig.\ref{3.pdf}, depending on the value of $U$, the realistic system can be either ferromagnetic or not. That is, for large enough $U$, in the realistic parameter regime, the ferromagnetic state is stable. Within the parameter regime used here, we estimate the critical value of $U_c$ to be around $U_c\approx 2 meV$.

Before ending this section, we'd like to emphasize the our formalism can be applied to \textit{arbitrary} interaction models, and that the model considered here only provides an example that can be generalized to more complicated models in future work.

\emph{Discussion and Conclusion}\label{c&d}- In this letter, we have shown that a simple variant of the ideas related to flat band ferromagnetism can be used to study the  in TBLG. In particular, we discussed ferromagnetism in the perfectly flat band ``chiral" limit, and used the exactly calculated one magnon spectrum to study the stability of the ferromagnetic state as the realistic system is approached. The same one-magnon spectrum is also used to extract the spin-stiffness of the ferromagnetic Goldstone modes. Note that our formalism can be used to study ferromagnetism in other recently discovered ferromagnetic phases in Moir\'e superlattices. In particular the same exact method can be readily applied to twisted double bilayer graphene where an analogous chiral flat limit exists\cite{vf3}.

A particularly intriguing feature of the results presented here is that (as opposed to mean-field approach\cite{mf1,mf2,1906.07302}) they manifestly predict ferromagnetism \textit{only} at odd filling fractions $\pm3/4,1/4$. Given that the experimentally observed half-filled state seems to be spin-unpolarized\cite{tbg1,tbg2,tbg3}, it would be interesting to study the fate of the model presented here at half-filling and to see if it also hosts a spin-unpolarized ground state.

The exactly calculated one-magnon spectrum studied here can be used to extract other interesting information about the ferromagnetic state. In particular, the Chern insulating nature of the ferromagnetic states means that the spin-stiffness can be used to calculate the energy of \text{charged} skyrmions\cite{skyrmion1,skyrmion2,skyrmion3,skyrmion4,skyrmion5,skyrmion6}. The skyrmion energy in combination with the Hartree-Fock particle hole excitation energy can then be used to determine whether skyrmions are the lowest lying charged excitations (note that even in Landau levels, this is not always the case). This results can be compared with experimentally measured charge gaps of Ref.\cite{young}.
We further remark that the one-magnon spectrum can be also used to identify natural candidates for the neighboring magnetic phases. Specifically, when the ferromagnetic state becomes unstable, that is when the minima of the one-magnon spectrum has finite momentum $q_0\ne 0$, the location of this new minima in the Brillouin zone can be used to identify natural candidates for alternate type of magnetic order (e.g. anti-ferromagnetism) that might replace ferromagnetism in neighboring phases.
 
We finally mention that the formalism developed here provides an intuitive picture of how ferromagnets are favored over competing Mott insulators in topologically non-trivial bands. Traditionally, when short-range Hubbard interactions are considered, Mott insulating states are as considered as candidate ground states. The idea is to restrict the electrons to sharply localized non-overlapping Wannier wave-functions to minimize the interaction energy. However, note that for non-isolated or isolated and topologically non-trivial bands, Wannier wave-functions are not even approximately localized. Therefore, in these cases (overlapping or topologically non-trivial band), Mott insulating states are not good ground state candidates. Whereas the ferromagnetic states discussed above, are good candidates independent of the (topological-)nature of the underlying band.  In continuation of these ideas, we mention here that recent experimental finding of Ref.\cite{tbg12} in ABC trilayer graphene, where the Chern number of the band can be electrically tuned, seems to suggest that the topology of the underlying band might in fact play a role in favoring ferromagnetism. Studying this case with the same formalism provides another interesting line of future work.

\section*{Acknowledgments}
 This work was supported by the NSF-DMR1555135 (CAREER), JQI-NSF-PFC (PHY1430094) and the Sloan research fellowship. We acknowledge the support of the University of Maryland supercomputing resources (http://hpcc.umd.edu/).
 
\emph{Note Added}--During the final stages of preparing this manuscript, we became aware of another preprint~\cite{1907.11723}. This paper also studies ferromagnetism in flat-band systems and has some overlap with the present work.

\bibliography{library}
\section{Supplementary Material}\label{sm}
\section{Formalism and Methods}\label{b0}
Here, we setup and introduce the a simple formulation of the general idea of ``flat-band ferromagnetism". We start with a heuristic discussion of ferromagnetism in the perfectly flat band limit. We then proceed to discussing the one-magnon excitation spectrum.

\subsection{Ferromagnetism in the perfectly flat band limit}\label{b1}
Consider a generic band system where the lowest occupied band is perfectly flat. We assume this band has an exact spin degeneracy. The flat band is spanned by a set of wave-functions, e.g. Bloch or Wannier wave-functions $\psi_m(x)$. We take the ($N_e$) particles in the lowest band to interact with a strictly contact, i.e. Hubbard, interaction,
\begin{align}
H_{\text{int}}= U \sum_{i<j=1}^{N_e} \delta(r_i-r_j).
\end{align}
Note that since the band is flat, the kinetic energy term is irrelevant, i.e. constant. Therefore, to find the ground state of the system we only need to consider the interaction term written above. Furthermore, note that the interaction term is non-negative $\langle H_{\text{int}}\rangle \ge 0$. 

Consider now a generic flat band wave function,
\begin{align}
\psi(x_1,x_2,...,x_{N_e}) \times | \psi_{\text{spin}} \rangle,
\end{align}
where the two parts represent spatial and spin (flavor) part of the the total wave-function. To minimize the energy and find the ground state, we need to ensure that $\psi(x_1,x_2,...,x_{N_e})$ vanishes when any two of coordinates are identical $x_i=x_j$. An easy (and perhaps the only generic) way to do this is to take the spatial wave wave-function $\psi(x_1,x_2,...,x_{N_e})$ to be totally anti-symmetric. However, since the entire wave-function also has to be anti-symmetric. The spin part of the wave function $|\psi_{\text{spin}} \rangle$ has to be totally \textit{symmetric}. It is then straightforward to show that \text{all} such wave-functions exhibit saturated ferromagnetism\footnote{To see this note that the permutation operator $\mathcal{P}_{ij}$ is related to the Heisenberg exchange operator.}. We emphasize that all this discussion also applies to lattice systems and is not restricted to the continuum.

 An important theorem due to Mielke\cite{mt2.5}, shows that if the one particle density matrix associated with the (Slater determinant) state discussed above is \text{irreducible} in real space, the saturated ferromagnetic states are the \textit{unique} ground states (up to trivial spin degeneracy). Conversely, if the density matrix is reducible, there are additional ground states corresponding to different $SU(2)$ rotations of different blocks of the density matrix. It is straightforward to show that in this case, the lowest energy states in the one-magnon (single spin flip) spectrum would be at least twofold degenerate. Throughout this work, and by direct calculation we have always ensured that the one-magnon spectrum is not degenerate. This puts additional context around the usefulness of the one-magnon spectrum as used in this paper. The generality and the simplicity of this results makes it a powerful tool in studying nearly flat band ferromagnetism. 

Traditionally, when short-range Hubbard interactions are considered, Mott insulating states are as considered as candidate ground states. The idea is to restrict the electrons to sharply localized non-overlapping Wannier wave-functions to minimize the interaction energy. However, note that for non-isolated or isolated and topologically non-trivial bands, Wannier wave-functions are not even approximately localized. Therefore, in these cases (overlapping or topologically non-trivial band), Mott insulating states are not good ground state candidates. Whereas the ferromagnetic states discussed above, are good candidates independent of the (topological-)nature of the underlying band.

We note that the argument above does not directly apply to the experimentally interesting case of TBLG. This is so because in that case there are additional subtleties caused by the extra degrees of freedom (valley, layer and sub-lattice) and the discrete symmetries associated with them (discussed in the main text). 

\subsection{The one-magnon spectrum and spin stiffness}
For all $SU(2)$ invariant systems, eigenstates can be classified with total spin $S_z$. The ferromagnetic manifold is constrained by the total spin. For the flat band limit discussed above, the ferromagnetic states are also the lowest energy states. A low energy set of excitations above the ferromagnetic states would be the states with slightly lower spin-polarization are obtained by flipping a single spin i.e. creating one magnon. The wave-functions in this manifold can be constrained to be of the form,
\begin{align} 
|\psi^q\rangle = \sum_{p\in \text{BZ}} \phi^q_p c^\dagger_{p+q,\downarrow} c_{p,\uparrow} |\uparrow \text{polarized} \rangle.
\end{align}
\begin{figure}[t]
\centering
	\vspace{1mm}
\includegraphics[width=0.9\columnwidth,keepaspectratio]{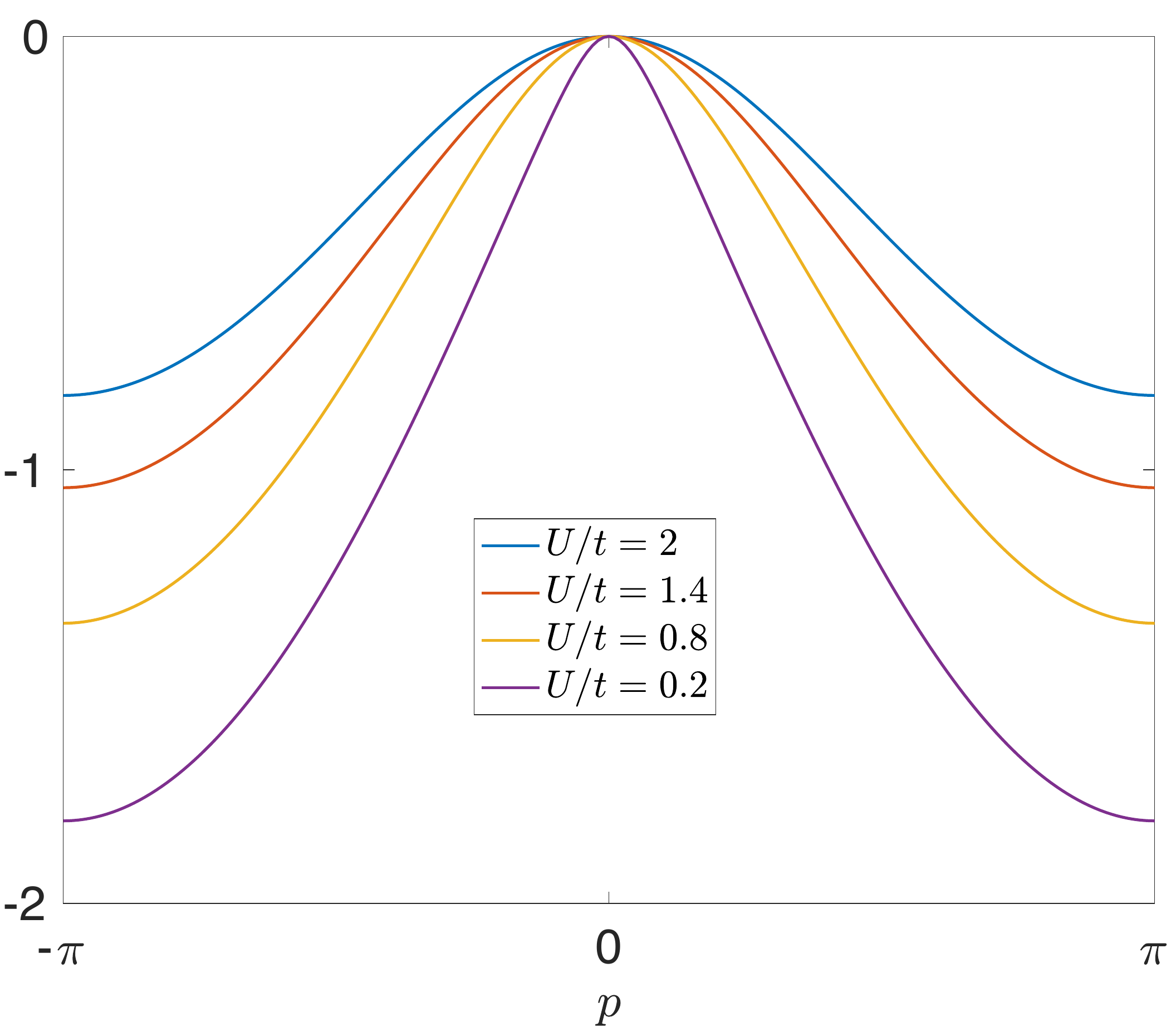} 
\caption{Lowest one-magnon band energy for the one dimensional Hubbard model in Eq.\eqref{hub}.}
\label{fig1}
\end{figure}
An identical relation also hold by changing $\uparrow \rightarrow \downarrow$ or by considering more general flavor degrees of freedom. 

As the ferromagnetic state becomes unstable, the one-magnon band minimum shifts from $q=0$ a finite momentum state $q\ne0$. It is then natural to consider a scenario where magnons condense at the location of the new minima, giving rise to a new type of magnetic order at finite momenta (e.g. anti-ferromagnetism). Therefore, in addition identifying the ferromagnetic region of the the phase diagram, we can use the one magnon spectrum to identify promising candidates for the neighboring phases. Note that the calculated one-magnon spectrum is not applicable to states with more than one magnon, and therefore, strictly speaking talking about magnon condensation is not meaningful in this picture. However, at least close to the ferromagnetic state, the same one-magnon spectrum is in principle calculable using the Holstein-Primakoff approximation. Within the Holstein-Primakoff approximation, the one-magnon spectrum is naturally extended to the entire many-body spectrum. Therefore, it is natural to justify magnon condensation scenarios from this point of view.

\section{Ferromagnetism and its stability in the one dimensional Hubbard model}\label{eh}

In this section we apply the machinery described so far the simple example of half-filled one dimensional Hubbard model,
\begin{align}\label{hub}
H=t(\sum_{i,\sigma} c^\dagger_{i,\sigma} c_{i+1,\sigma} + h.c.) + U \sum_i n_{i,\uparrow} n_{i,\downarrow}
\end{align}
The perfectly flat band limit in this case corresponds to the somewhat pathological case of $t=0$. In this case, the system has a large $2^N$ dimensional grounsntate subspace (including the fully saturated ferromagnetic state) corresponding to filling only electron per-site. In addition, we know that as the bandwidth $t$ is increased, the ground state becomes anti-ferromagnetic. In fact this system does not exhibit ferromagnetism anywhere in its phase diagram. 

To demonstrate validity, it is important to show that we can get the same results with our approach. As described earlier, we do so by calculating the one-magnon spectrum of this model. That is, we consider states of the form,
\begin{align} 
|\psi^q\rangle = \sum_p \phi^q_p c^\dagger_{p+q,\downarrow} c^\dagger_{p,\uparrow} |\uparrow \uparrow ... \uparrow  \rangle,
\end{align}
and calculate the lowest-band of energies associated with such excitations. Results of this calculation are shown in Fig.\ref{fig1}. As shown in Fig.\ref{fig1}, the ferromagnetic state $q=0$, is in fact a local maxima for all values of $U/t$. That is, ferromagnetism is always unstable (spin-stiffness is negative), and therefore, the true ground-state of the system is never ferromagnetic. This results shows the effectiveness of our method in ruling out ferromagnetism in one dimensional Hubbard model.
\begin{figure}[t]
\centering
	\vspace{1mm}
\includegraphics[width=\columnwidth,keepaspectratio]{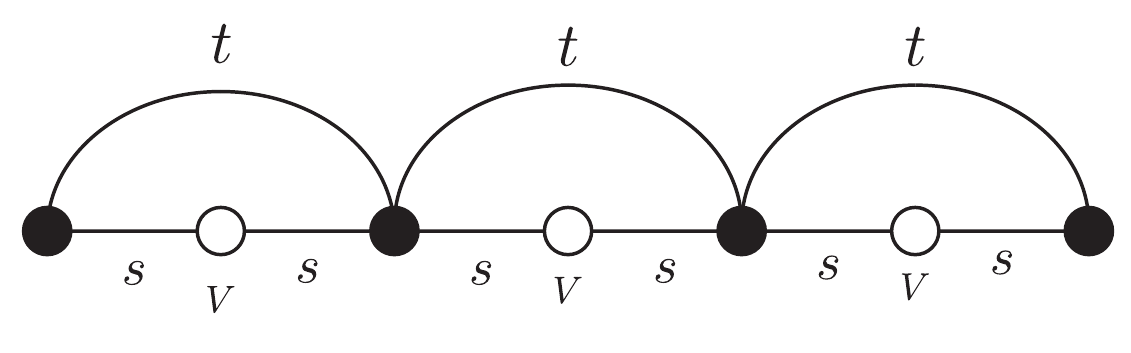} 
\caption{Hopping parameters in the Tasaki model Eq.\eqref{tski}.}
\label{fig2}
\end{figure}

\section{Ferromagnetism and its stability in the one dimensional Tasaki model}\label{et}

We now consider the application of our formalism to the one dimensional model of Tasaki\cite{tasaki},
\begin{align}\label{tski}
H&=t\sum_{i,\sigma} (c^\dagger_{2i,\sigma} c_{2i+2,\sigma} + h.c.) + s\sum_{i,\sigma} (c^\dagger_{i,\sigma} c_{i+1,\sigma} + h.c.)  \\ \newline\nonumber
&+  V \sum_{i,\sigma} n_{2i+1,\sigma} + U \sum_i n_{i,\uparrow} n_{i,\downarrow}
\end{align}

This model provides the simplest example of \text{stable} flat band ferromagnetism in one-dimension. A pictorial representation of the model and model and hopping parameters is shown in Fig.\ref{fig2} . Note that since the unit cell is composed of two-sites, this model has two bands. 

It is convenient to introduce auxiliary parameters $\lambda$ and $\rho$,
\begin{figure}[t]
\centering
	\vspace{0.5mm}
\includegraphics[width=\columnwidth,keepaspectratio]{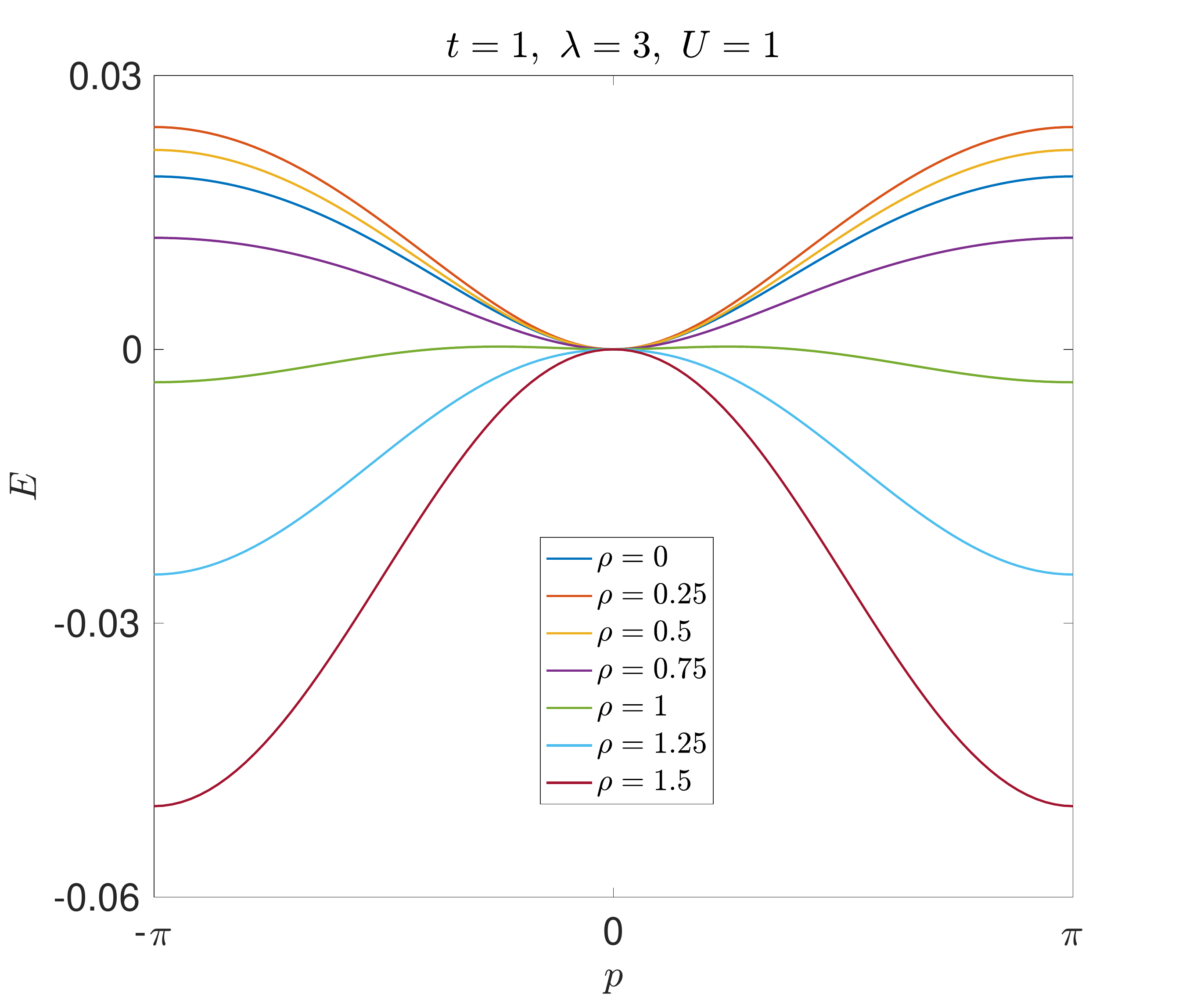} 
\caption{Lowest one-magnon band energy for the one dimensional Tasaki model in Eq.\eqref{tski}.}
\label{fig3}
\end{figure}
\begin{align}
s=\lambda t ~~;~~ V= (\lambda^2-2+\rho)t.
\end{align}
If we set $\rho=0$, the lowest band becomes perfectly flat. Roughly speaking, the parameters $|\rho| t$ and $\lambda^2 t$ set the band width and the band gap of the spectrum. In the flat band limit $\rho=0$ and at half filling, the ground state is a saturated ferromagnet. As the bandwidth is increased, the (saturated) ferromagnetic phase persists for a finite region. In the opposite limit where the the interaction strength $U$ is smaller than the bandwidth, the ground state is know to be a total spin singlet\cite{tasaki}. Therefore, as the bandwidth is increased from zero, the system goes from a ferromagnet to a singlet state.

 Similar to the Hubbard model, we calculate the one magnon spectrum as the band-width is increased. The calculated one-magnon spectrum is plotted is Fig.\ref{fig3}. As shown in the figure, the ferromagnetic states goes from a local minima (stable ferromagnetism) to a local maxima (unstable ferromagnetism) as the bandwidth is increased. Our results match the known analytic expression of spin-stiffness in the large band gap ($\lambda^2 t$) limit\cite{tasaki}.
\end{document}